\documentstyle[12pt]{article}
\newcommand{\AmS}{{\protect\the\textfont2
  A\kern-.1667em\lower.5ex\hbox{M}\kern-.125emS}}
\newcommand{\nc}{\newcommand}
\nc{\be}{\begin{displaymath}}
\nc{\ee}{\end{displaymath}}
\nc{\bea}{\begin{eqnarray*}}
\nc{\eea}{\end{eqnarray*}}
\nc{\rbo}{\raisebox}
\nc{\cH}{{\cal H}}
\nc{\RR} {\rangle \! \rangle}
\nc{\LL} {\langle \! \langle}
\nc{\rmi}[1]{{\mbox{\small #1}}}
\nc{\eq}{eq.~}
\nc{\nr}[1]{(\ref{#1})}
\nc{\ul}{\underline}
\nc{\cM}{{\cal M}}
\nc{\mc}{\multicolumn}
\nc{\todo}[1]{\par\noindent{\bf $\rightarrow$ #1}}

\nc{\nonu}{\nonumber}
\nc{\lmax}{l_{\rm max}}
\nc{\half}{\mbox{\small$\frac12$}}
\nc{\eights}{\mbox{\small$\frac18$}}
\nc{\bg}{\bar \gamma}

\nc{\hga}{{\widehat \gamma }}
\nc{\VVert}{{|\! \! \; |\! \! \; | }}
\nc{\twtimes}{{\widetilde \times }}
\nc{\twz}{{\widetilde z }}
\nc{\lst}{{l^*}}
\nc{\dst}{{d^*}}
\nc{\bC}{{\bf C}}
\nc{\bN}{{\bf N}}
\nc{\bP}{{\bf P}}
\nc{\bR}{{\bf R}}
\nc{\bZ}{{\bf Z}}
\nc{\cA}{{\cal A}}
\nc{\cB}{{\cal B}}
\nc{\cC}{{\cal C}}
\nc{\cF}{{\cal F}}
\nc{\cG}{{\cal G}}
\nc{\cL}{{\cal L}}
\nc{\cN}{{\cal N}}
\nc{\cP}{{\cal P}}
\nc{\cR}{{\cal R}}
\nc{\cS}{{\cal S}}
\nc{\cT}{{\cal T}}
\nc{\cU}{{\cal U}}
\nc{\cZ}{{\cal Z}}
\nc{\om}{{\overline m}}
\nc{\oU}{{\overline U}}
\nc{\oX}{{\overline X}}
\nc{\g}{{\gamma}}
\nc{\lam}{{\lambda}}
\nc{\Lam}{{\Lambda}}
\nc{\rv}{{\rm v}}
\nc{\eins}{{\bf 1}}
\nc{\Gau}[2]{d\mu_{#1}({#2})}
\nc{\iGau}[2]{\int d\mu_{#1}({#2})}
\nc{\Gaug}{d\mu_{\g}({\Phi })}
\nc{\Gaurv}{d\mu_{\rv }({\Phi })}
\nc{\iGaug}{\int d\mu_{\g}({\Phi })}
\nc{\GauN}[2]{d\mu_{#1}^{(N)}({#2})}
\nc{\iGauN}[2]{\int d\mu_{#1}^{(N)}({#2})}
\nc{\iGaugN}{\int d\mu_{\g}^{(N)}({\Phi })}
\nc{\GaugN}{d\mu_{\g}^{(N)}({\Phi })}
\nc{\GaurvN}{d\mu_{\rv }^{(N)}({\Phi })}
\nc{\Ap}{{A^{\prime }}}
\nc{\Lp}{{\Lambda^{\prime }}}
\nc{\Vp}{{V^{\prime }}}
\nc{\Zp}{{Z^{\prime }}}
\nc{\up}{{u^{\prime }}}
\nc{\yp}{{y^{\prime }}}
\nc{\zp}{{z^{\prime }}}
\nc{\inn}{{\underline \in }}

\title{Hierarchical renormalization group fixed points}
\author{A. Pordt\thanks{Work supported by the Deutsche Forschungsgemeinschaft
                        under Grant Wi 1280/2-1} \\Institut
        f\"ur Theoretische Physik I,
	Universit\"at M\"unster, \\
	Wilhelm-Klemm-Str. 9, D-48149 M\"unster, Germany}

\begin{document}
\maketitle

\begin{abstract}
Hierarchical renormalization group transformations are related to
non-associative algebras. Non-trivial infrared fixed points are
shown to be solutions of polynomial equations. At the example
of a scalar model in $d(\ge2)$ dimensions some methods for
the solution of these algebraic equations are presented.
\end{abstract}


\section{Introduction}

Wilson's renormalization group (RG) approach \cite{Wi74}
is an important tool for the
investigation of quantum field theories. The study of the ultraviolet
(UV) and infrared (IR) limits correspond to the study of field theories near
fixed points of a renormalization group transformation. A trivial
fixed point (UV fixed point) corresponds to asymptotically free theories.
For these models standard perturbation methods are sufficient. There
are other non-trivial RG fixed points (IR fixed points)
where a small parameter is lacking. The goal of this paper is to
present non-perturbative methods for analyzing models at IR fixed
points.

Investigation of quantum field theories at infrared fixed points
requires non-perturbative methods. In this paper we want to consider
hierarchical RG fixed points.

The hierarchical RG
transformation (HRGT) for a scalar field theory is defined in the following.
Let $Z : \bR \rightarrow \bR $ be a (Gaussian-measurable) function
(which corresponds to the generating functional of free
propagator-amputated Greens functions).
$Z$ is also called {\em partition function}.
Application of the HRGT $\cR \cG $ to the partition
function $Z$ yields the {\em effective partition function} $\cR \cG (Z) :
\bR \rightarrow \bR $ defined by
      \begin{equation} \label{HRGT}
        \cR \cG (Z)(\Psi ) := \iGaug \left[
         Z(\Phi +L^{1-\frac{d}{2}}\Psi ) \right]^{L^d},
      \end{equation}
where $\Psi \in \bR $, $L \in \{ 2,3,\ldots \} $, $\gamma >0$ and
the Gaussian measure is defined by
      \begin{equation}
        \iGaug := \int_{-\infty}^\infty d\Phi \exp \{ -
          \frac{\Phi^2}{2\gamma} \} .
      \end{equation}
The factor $L$ corresponds to the block size of a block spin RG
transformation and $\g $ corresponds to the fluctuation propagator.
The generalization to $N$-component
models is straightforward.

We want to find even fixed points $Z^*$ of the HRGT $Z^* = \cR \cG (Z^*).$
Excluding $Z^* = 0$ or $\infty$, we immediately find the following
trivial fixed points. There are the UV fixed point $Z_{UV}(\Psi )=1$ and
the high-temperature fixed point $Z_{HT}(\Psi ) = L^{\frac{1}{L^d-1}}
\exp \{ -\frac{L^2-1}{2\g L^d}\Psi^2 \} .$ For $d\ge 4$ there are only these
trivial fixed points. In $2\le d<4$ dimensions it turns out that there
exists also other non-trivial fixed points which are called infrared
fixed points. For a further investigation one introduces the hierarchical
RG algebra \cite{PoWi94}.

\section{Hierarchical renormalization group algebras}

Let $P(\Phi )$ be a polynomial in $\Phi .$ Define normal-ordering
(Wick-ordering) of $P(\Phi )$ by
\begin{equation}
:P(\Phi ):_\g := \exp \{ - \frac{\g}{2} \frac{\partial^2}{\partial \Phi^2} \}
 \, P(\Phi ).
\end{equation}
Introduce structure coefficients $\cC_l^{mn}$ by the fusion relations
      \begin{eqnarray}
      \lefteqn{:\Phi^{2m}:_\g \cdot :\Phi^{2n}:_\g =}\\ & &
       \sum_{l:\, |m-n| \le l\le m+n}
       \cC_l^{mn} :\Phi^{2l}:_\g .
      \end{eqnarray}
Let $\{ e_m,m = 0,1,2,\ldots \} $ be the canonical basis of the vector space
$\bR^\infty .$ Define a $\times $-product on $\bR^\infty $ by
      \begin{equation}
      e_m \times e_n :=
       \sum_{l:\, |m-n| \le l\le m+n}
       \cC_l^{mn} e_l.
      \end{equation}
Then $(\bR^\infty ,\times )$ is a commutative and associative algebra with
unit element $e_0.$

We will identify partition functions $Z :\bR \rightarrow
\bR $ with elements $z=(z_0,z_1,z_2,\ldots )=\sum_{n=0}^\infty z_ne_n$
of the algebra $\bR^\infty $ by
      \begin{equation}
        Z(\Phi ) = \sum_{n=0}^\infty \frac{z_n}{\gamma^n}
        :(\Phi^{2})^n:_\gamma .
      \end{equation}
For $z\in \bR^\infty $ and $\beta \in \bR $ define the scaling relation
$\cS_\beta :\bR^\infty \rightarrow \bR^\infty $ by $\cS_\beta (z)_l :=
\beta^{2l} z_l.$ Then one can show that the HRGT eq.~(\ref{HRGT}) -- where
$\gamma $ is replaced by $(1-\beta^2)\gamma $ -- is equivalent to
      \begin{equation}
        \cR \cG (z) = \cS_\beta (
          \underbrace{z\times \cdots \times z}_{L^d\, factors}),
      \end{equation}
where $\beta := L^{1-\frac{d}{2}}.$ Let us remark that this relation
is only helpful if $|\beta | < 1.$ This seems to exclude the case $d=2.$
We may avoid this problem by a transformation of the HRGT eq.~(\ref{HRGT}).
Define a new partition function $F$ by $F := Z/Z_{HT}.$ Then the HRGT
of $F$ has the same form as the HRGT eq.~(\ref{HRGT}) of $Z.$ Only the
coefficient $\g $ has to be replaced by $L^{-2}\g $ and the scaling
parameter $\beta$ by $L^{-2}\beta = L^{-1-\frac{d}{2}} <1.$
This method is called ``extraction of the high-temperature fixed point''.

For the simplest choice $L^d=2$ the fixed points $z^*$ are solutions
of the quadratic equation $z = z\times_\beta  z:= \cS_\beta (z\times z).$
The product $\times_\beta $, for $\beta \ne 1,$ defines a commutative
and non-associative
algebra $(\bR^\infty ,\times_\beta )$ containing no unit element. The algebra
$(\bR^\infty ,\times_\beta )$ is called a {\em hierarchical renormalization
group algebra} (HRGA). By change of coordinates there are
also other HRGA's possible.
For example, identify the partition functions $Z :\bR \rightarrow
\bR $ with elements $\twz =(\twz_0,\twz_1,\twz_2,\ldots )=
\sum_{n=0}^\infty \twz_ne_n$
of the algebra $\bR^\infty $ by series expansion
      \begin{equation}
        Z(\Phi ) = \sum_{n=0}^\infty \frac{\twz_n}{\gamma^n}
        (\Phi^{2})^n .
      \end{equation}
Then one can show that the HRGT eq.~(\ref{HRGT}) is
equivalent to
      \begin{equation}
        \cR \cG (\twz) = \cS_\beta (
          \twz \star \twz ) = \twz \star_\beta \twz ,
      \end{equation}
where the $\star_\beta $-product defines the non-associative algebra
$(\bR^\infty ,\star_\beta )$ and the corresponding structure coefficients
$\cS_l^{mn}$ can be easily computed \cite{PiPoWi94,PoWi94}. Let us remark
that this product cannot be used for cases $L^d \ne 2,$ since
$(\bR^\infty ,\star_\beta )$ is also non-associative for $\beta = 1.$

For periodic fixed points in $d=2$ dimensions one represents the partition
function $Z$ with period $T=\frac{\sqrt{\g}}{q}$ by
      \begin{equation}
        Z(\Phi ) = \sum_{m 0 -\infty}^{\infty} z_m
           \exp \{ 2\pi im\frac{q}{\sqrt{\gamma}} \Phi \}
      \end{equation}
Define a $\circ_q$-product by
      \begin{equation}
        e_m \circ_q e_n := \exp \{ -2\pi^2 q^2 (m+n)^2\} e_{m+n}.
      \end{equation}
Then the HRGT eq.~(\ref{HRGT}), for $d=2$ and $L^2=2,$ is equivalent to
      \begin{equation}
        \cR \cG (z) = z\circ_q z
      \end{equation}
The HRGA $(\bR^\infty ,\circ_q)$ is non-associative for $q\ne 0.$

\subsection{Norms}

For estimations and convergence proofs suitable definitions of norms for
HRGA's are required.
Equipped with these norms the algebras $\bR^\infty$ become Banach-algebras.

The norms $\Vert \cdot \Vert_\rho ,$ defined below,
depend on a parameter $\rho .$ If the parameter $\rho $ is large
enough, $\rho \ge \rho^*,$ the norms become algebra-norms, i.~e. they
obey
      \begin{equation}
        \Vert a \bullet b\Vert_\rho \le \Vert a\Vert_\rho \cdot
          \Vert b\Vert_\rho ,
      \end{equation}
where the $\bullet$-product defines a HRGA. For example, Koch and Wittwer
\cite{KoWi86} used the following norm for their proof of a non-trivial
fixed point in $d=3$ dimensions for the case $L^d=2$ and extracting the
high-temperature fixed point $(2\beta^2 <1)$
      \begin{equation}
        \Vert z\Vert_\rho^{(1)} := \sum_{n=0}^\infty \sqrt{(2n)!} |z_n|
            \rho^n .
      \end{equation}
For the use of estimating coefficients in the $\epsilon$-expansion
the following sup-norm is useful \cite{PoWi94} $(2\beta^2 >1)$
      \begin{equation}
        \Vert z\Vert_\rho^{(\infty )} := \sup_{n}( n! |z_n|\rho^n ).
      \end{equation}
In constructive field theory norms are typically defined by a large
field regulator $e^{\rho \Phi^2}$
       \begin{equation}
        \Vert z\Vert_\rho^{(c)} := \sup_{\Phi \in \bR}
         |e^{\rho \Phi^2} Z(\Phi )|.
        \end{equation}

\section{Methods and Results}

There are several methods to solve the fixed point equation.
Beside the method of $\epsilon$-expansion all methods presented here
deliver convergent expansions for the fixed points.

The first and most
standard one is to iterate the RGT. Define a sequence of partition
functions by $z^{(n+1)} := \cR \cG (z^{(n)}).$ One has to choose the
start of the RGT $z^{(0)}$ such that $\lim_{n\rightarrow \infty} z^{(n)}=
z^*$ exists. Then $z^*$ is a fixed point.

A second standard method is the $\epsilon$-expansion
\cite{CoEc77,PiPoWi94}. Take the trivial UV fixed point
$z_{UV} =e_0$ in $d_*$ dimensions. Then one considers the fixed point
equation in $d = d_* -\epsilon $ dimensions. The fixed point $z^*$ and
the scaling factor $\beta^{2l}$ is written as a power series in $\epsilon $.
Insertion of these series in the fixed point equation $z= z\times_\beta z$
and comparison of the coefficients yields recursive equations for the
coefficients of the fixed point in the case of critical dimensions
$d_* = d_l :=\frac{2l}{l-1}$, $l\in \{ 2,3,\ldots \} .$
$\ln Z^* (\Phi )$ has the form of an $l$-well.

That $d_l$ are the
critical dimensions can be seen in the following way. We suppose that for
small $\epsilon $ the fixed point $z = e_0 +r$ is near by the UV fixed point
$e_0$, i.~e. $r$ is small. The fixed point equation in terms of $r$ is
$r = 2e_0 \times_\beta r + r\times_\beta r.$ In lowest order $r$ should
obey $r\in \ker (id - 2e_0\times_\beta )$. The linear operator
$(id - 2e_0\times_\beta ) = diag (1-2\beta^{2l},l=0,1,\ldots )$ has only
a non-trivial kernel if there exists $l$ such that $1 = 2\beta^{2l}.$
This is the case for the critical dimensions $d=d_l.$

A third method used by Koch and Wittwer \cite{KoWi86}
is the method of exact beta-functions. For that, one defines a
projection operator ${\rm P}$ on $\bR^\infty $ such that
      \begin{equation}
        z^{rel} = z^{rel}(\g_0,\ldots ,\g_N) = {\rm P}(z)
      \end{equation}
depends only on a finite number of parameters $\g_0,\ldots ,\g_N$
and such that the irrelevant part of $z$ defined by $r := (1-{\rm P})(z)$
can be controlled by standard fixed point theorems. The irrelevant
fixed point $r^* = r^*(\g )$ is defined by
      \begin{equation}
        r^* = H_{\g }(r^*) := (\eins -{\rm P})\cR \cG (z^{rel}(\g ) + r^*).
      \end{equation}
The exact beta-function $B : \bR^{N+1} \rightarrow \bR^{N+1}$ is defined
by
      \begin{eqnarray}
        \lefteqn{z^{rel} (B(\g_0,\ldots ,\g_N)) =} \\
          & & {\rm P} \cR \cG \left(
           z^{rel}(\g_0,\ldots ,\g_N) +
           r^* (\g_0,\ldots ,\g_N)\right) .
      \end{eqnarray}
The fixed points $z^* =\cR \cG (z^*)$ are given by
$z^* = z^{rel}(\g^*) + r^* (\g^*),$ where $\g^*$ is the fixed point of
the exact beta-function $B$, $B(\g^*) = \g^*.$ For the HRGT one can use
the projection operator
      \begin{equation}
        {\rm P}(z) := (z_0,\ldots ,z_N,0,0,\ldots )
      \end{equation}
and the beta-function technique works for $N\ge N_0 = O(1) (=7).$

A fourth method is to replace the fixed
point equation by a minimization-problem. Define an energy-function
$E(z)$ for $z\in \bR^\infty $ by
      \begin{equation}
        E(z) := \sum_{l=0}^\infty (z_l - (z\times_\beta z)_l)^2 \ge 0.
      \end{equation}
Then $z=z\times_\beta z$ is equivalent to $E(z)=0.$

A fifth method is to solve the equation $z\times_\beta z - z=0$ directly
by Newton's method \cite{PiPoWi94}.

A sixth method is to truncate the infinite system of quadratic
equations by setting $z_l=0$ if $l>N$ and solve the finite system
directly by means of Groebner bases.

For convergence proofs and numerical calculations one has
to truncate the infinite system.
One can show that there exists a solution of the truncated fixed point
equation $z^{(N)}$ such that the limit $\lim_{N\rightarrow \infty}z^{(N)}
=z^*$ exists.

\end{document}